%% file: distr_triangle_2D.tex
\documentclass[sigconf,9pt]{acmart}

\def\BibTeX{{\rm B\kern-.05em{\sc i\kern-.025em b}\kern-.08emT\kern-.1667em\lower.7ex\hbox{E}\kern-.125emX}}
    
\copyrightyear{2019} 
\acmYear{2019} 
\setcopyright{acmcopyright}
\acmConference[ICPP 2019]{48th International Conference on Parallel Processing}{August 5--8, 2019}{Kyoto, Japan}
\acmBooktitle{48th International Conference on Parallel Processing (ICPP 2019), August 5--8, 2019, Kyoto, Japan}
\acmPrice{15.00}
\acmDOI{10.1145/3337821.3337853}
\acmISBN{978-1-4503-6295-5/19/08}

\usepackage{amsmath,amssymb,amsfonts}
\usepackage{algorithmic}
\usepackage{array}
\usepackage{textcomp}
\usepackage{amssymb}
\usepackage{graphicx}
\usepackage{url}
\usepackage[flushleft]{threeparttable}
\usepackage{booktabs}
\usepackage{amsopn}

\usepackage{mathtools}

\usepackage{graphicx}  
\usepackage[caption=false]{subfig}

\usepackage{subfig}
\usepackage{graphicx}
\usepackage{tikz}

\DeclareMathOperator{\Adj}{Adj}
\DeclareMathOperator{\AdjG}{\mbox{Adj$^>$}}
\DeclareMathOperator{\AdjL}{\mbox{Adj$^<$}}

\def\BibTeX{{\rm B\kern-.05em{\sc i\kern-.025em b}\kern-.08em
    T\kern-.1667em\lower.7ex\hbox{E}\kern-.125emX}}

\usepackage{array}
\newcolumntype{L}[1]{>{\raggedright\let\newline\\\arraybackslash\hspace{0pt}}m{#1}}
\newcolumntype{C}[1]{>{\centering\let\newline\\\arraybackslash\hspace{0pt}}m{#1}}
\newcolumntype{R}[1]{>{\raggedleft\let\newline\\\arraybackslash\hspace{0pt}}m{#1}}

\begin{document}

\title{A 2D Parallel Triangle Counting Algorithm for Distributed-Memory Architectures}

%

\author{Ancy Sarah Tom and George Karypis}
\affiliation{%
 \institution{Department of Computer Science and Engineering, University of Minnesota}
}
\email{tomxx030@umn.edu, karypis@cs.umn.edu}

%
\renewcommand{\shortauthors}{Tom and Karypis}

%
\begin{abstract}
Triangle counting is a fundamental graph analytic operation that is used extensively 
in network science and graph mining. As the size of the graphs that 
needs to be analyzed continues to grow, there is a requirement in developing 
scalable algorithms for distributed-memory parallel systems.  
%
To this end, we present a distributed-memory triangle counting algorithm,
which uses a 2D cyclic decomposition to balance the computations and 
reduce the communication overheads.
%
%
The algorithm structures its communication and computational
steps such that it reduces its memory overhead and
includes key optimizations that leverage the sparsity of 
the graph and the way the computations are structured.
%
%
Experiments on synthetic and real-world graphs show that our algorithm obtains an average relative speedup range between $3.24$ to $7.22$ out of $10.56$ across the 
datasets using $169$ MPI ranks over the performance achieved by $16$ MPI ranks. 
Moreover, we obtain an average speedup of $10.2$ times on comparison with 
previously developed distributed-memory parallel algorithms.
\end{abstract}

%
%


%

%
\maketitle

\input{intro}

\input{defs}

\input{background}

\input{related}

\input{methods}

\input{experimental}

\vspace*{-3mm}
\input{conclusion}

%
\begin{acks}
This work was supported in part by NSF (1447788, 1704074, 1757916, 1834251), Army
Research Office (W911NF1810344), Intel Corp, and the Digital Technology Center at the
University of Minnesota. Access to research and computing facilities was provided by
the Digital Technology Center and the Minnesota Supercomputing Institute.
\end{acks}

%
\bibliographystyle{ACM-Reference-Format}
\bibliography{distr_triangle_2D}

\end{document}

%% file: intro.tex
\section{Introduction}
\label{intro}

The use of graphs to model large scale real-world data is ubiquitous in our everyday lives. 
In order to analyze and study the relationships these graphs 
model, graph analytic operations such as 
finding patterns of interest, analyzing the community structure and connectivity, and 
determining influential entities in a given graph, 
are commonly used. 

One such graph analytic operation 
is counting the number of triangles in a graph. The number of triangles 
in a graph is an important statistic that is used as an intermediary step in various applications. 
It is used in computing the clustering coefficient and the transitivity ratio 
of graphs, both of which are used in characterizing the tendency of the 
nodes in a graph to cluster together. 
Furthermore, the computations involved in triangle counting forms an important step 
in computing the $k$-truss 
decomposition of a graph, detecting community 
structures, studying motif occurrences, detecting spamming 
activities, and understanding the structure of biological networks~\cite{girvan2002community,smith2017truss,shrivastava2008mining,watts1998collective}.
 
In recent years, driven by the growing size of the graphs that needs to be analyzed, there has been 
significant research in improving the efficiency of parallel algorithms for computing 
the exact and approximate number of triangles. Parallel triangle counting algorithms 
have been specifically built for GPUs, external memory, shared-memory, and distributed-memory 
platforms~\cite{green2014fast,hu2018high,bisson2018update,polak2016counting,wang2016comparative,arifuzzaman2017distributed,parimalarangan2016fastpaper,shun2015multicore,yacsar2018fast}. 
The shared-memory class of solutions are limited by the amount of memory that is available in a 
single processor, thus, limiting the size of the graphs that can be analyzed. 
Moreover, in many practical settings, such large graphs are stored in a 
distributed fashion in the aggregate memory that is available in a 
distributed-memory system. Being able to successfully analyze large graphs 
in such scenarios
requires the need of developing distributed-memory algorithms for counting the 
number of triangles.
However, despite the advantages of a distributed-memory system, these algorithms 
face higher costs as compared to shared-memory systems during communication 
and synchronization steps. Furthermore, distributed-memory 
graph algorithms also entail the problem of intelligent graph partitioning 
in order to reduce the costs involved in communicating with 
neighboring vertices.

We present an MPI-based distributed-memory algorithm for triangle counting using a set intersection based approach~\cite{tom2017exploring}. The key difference between our algorithm and previously proposed approaches is that it utilizes a 2D decomposition of the data and associated computations, which increases the concurrency that can be exploited and reduces the overall communication cost. Furthermore, our algorithm moves the data among the processors by utilizing a sequence of communication steps that are similar to those used by Cannon's parallel matrix-matrix multiplication algorithm. This ensures that our algorithm is memory scalable 
and faces low communication overhead. 
%
We also include key optimizations that leverage the sparsity of the graph and the way the computations are 
structured. Some of these optimizations include enumerating the triangles using an ordering that specifically leverages hash-maps in the set intersection computation, changing the hashing routine for vertices based on the density of its adjacency list, and eliminating unnecessary intersection operations. 

We evaluate the performance of our algorithm on various real-world and synthetically generated graphs and compare it against other existing state-of-the-art approaches. Our experiments show that 
we obtain a relative speedup that range between $3.24$ and $7.22$ out of $10.56$
across the datasets using 169 MPI ranks over the performance achieved by 16 MPI ranks. 
Moreover, the performance of our parallel algorithm compares favorably against those achieved by existing distributed memory algorithms that rely on 1D decomposition~\cite{6569865,arifuzzaman2017distributed, kanewala2018distributed}.

The rest of the paper is organized as follows. Section~\ref{defs} introduces the notation and definitions used in the paper, Section~\ref{background} details the necessary background required, 
Section~\ref{related} describes competing approaches we use, Section~\ref{methods} 
explains our 2D parallel algorithm and finally, Section~\ref{experimental} details
several experiments which demonstrate our algorithm's speedup and scaling capabilities. 

%% file: defs.tex
\section{Definitions and notations}
\label{defs}

\noindent
We will assume that the graphs that we operate on are \emph{simple} and
\emph{undirected} and are represented using the standard $G=(V, E)$ notation.
We will use $\Adj(v_i)$ to denote the \emph{adjacency list} of $v_i$; i.e., the set of
vertices that are adjacent to $v_i$.
We will use $d(v_i)$ to denote the \emph{degree} of $v_i$, i.e., $d(v_i) =
|\Adj(v_i)|$. We will use $\%$ to indicate a mod operation and $\div$ to indicate
a divide operation.

We will use $A$ to denote the $n\times n$ adjacency matrix of a symmetric 
$n$ vertex graph, in which $a_{i,j} = 1$, if there is an edge 
between $v_i$ and $v_j$, and 0, otherwise. Furthermore, we will use 
$U$ and $L$ to denote the upper and the lower triangular 
portion of the adjacency matrix of $G$. 
A \emph{triangle} is a set of three vertices $\{v_i, v_j, v_k\}$ if the edges $(v_i,
v_j)$, $(v_i, v_k)$, and $(v_j, v_k)$ exist in $E$.
The problem of \emph{triangle counting} is to compute the total number of unique
triangles in $G$.

Lastly, let there be $p$ processors in the system, which can be 
arranged in a square grid of the form $\sqrt{p}\times\sqrt{p}$. The processor in the $i$th 
row and $j$th column in this square grid is denoted by $P_{i,j}$.


%% file: background.tex
\section{Background}
\label{background}
\subsection{Triangle counting}
%
Triangle counting algorithms iterate over all the edges in a graph and for each edge, count the number of triangles that this edge is a part of. To ensure that they do not count the same triangle multiple times, they impose strict vertex enumeration rules and ignore any triangles that do not satisfy those. There are two such rules, which are referred to as $\langle i, j, k\rangle$ and $\langle j, i, k\rangle$, where $i<j<k$. The $\langle i, j, k\rangle$ rule dictates that the algorithm starts from the first vertex ($i=1$) and iterates over the non-zero entries of $U$ in a row-wise fashion (i.e., edges of the form $(v_i, v_j)$) and considers only the triangle-closing vertices $v_k$, where $j<k$. In contrast, the $\langle j, i, k\rangle$ rule dictates that the algorithm also starts from the first vertex ($i=1$) and considers the triangle-closing vertices $v_k$, where $j<k$, but, iterates over the non-zero entries of $U$ in a \emph{column-wise fashion} (i.e., edges of the form $(v_j, v_i)$).

Using the $\langle i, j, k\rangle$ rule, it is easy to show that the 
number of triangles, $c_{i, j}$, that result from $U$'s $(v_i, v_j)$ edge is
%
%
\begin{equation} \label{eq:1}
  {c_{i,j}} = \sum\limits_{k=j+1}^{n} u_{i,k} \times u_{j,k},
\end{equation}
\noindent
and the total number of triangles in the graph is 
\begin{equation} \label{eq:2}
  \sum_{a_{i,j}=1 \land  i<j} c_{i,j}.   
\end{equation}
\noindent

%

The computation associated with $c_{i,j}$ can be viewed as a set-intersection 
operation between the adjacency lists of $v_i$ and $v_j$, where $i<j<k$, and 
can be modeled as a list-based intersection and a map-based intersection. 
For every edge $(v_i,v_j)$ in $U$, the list-based intersection involves jointly traversing 
the adjacency lists of vertex $v_i$ and $v_j$, and finding common vertices, $v_k$, 
between them, such that $j<k$. In the map-based approach, we use an auxiliary data structure and hash 
$U_{i,*}$ (or, $U_{j,*}$) within that. This auxiliary data structure is then used to lookup 
the vertices of $U_{j,*}$ (or, $U_{i,*}$). Each successful lookup accounts 
for a complete triangle. 
On comparing the performance of the two approaches, we 
observe that map-based approaches are faster
than list-based approaches. This is because for each $v_i$, the 
hash-map used to store the adjacency list of $v_i$ can be reused for
all $v_j \in U_{i,*}$. The list-based and map-based methods are further detailed in~\cite{parimalarangan2016fastpaper,shun2015multicore,tom2017exploring}. 

Note that an alternative way of writing Equation~\ref{eq:1} is 
\begin{equation} \label{eq:6}
c_{i,j} = \sum_{k=1}^{n} u_{i,k}\times l_{k,j},
\end{equation}
since all entries of $L$ for $k<j$ are by definition 0. Given this, we can conceptually view the computations associated with triangle counting as that of computing certain entries of the matrix that results from multiplying matrices $U$ and $L $. Specifically, the entries that we care 
about are those that correspond to the non-zero entries of $U$. 
We will use $C[U]$ to denote the non-zero entries of U, and we will use 
\begin{equation} \label{eq:7}
C[U] = UL
\end{equation} 
to denote this operation of computing the entries of $C[U]$.

Furthermore, studies~\cite{arifuzzaman2017distributed} have shown that ordering 
the vertices in non-decreasing degree before the triangle counting step leads to lower runtimes. 
%
%
%
This optimization has been incorporated in most triangle 
counting algorithms that have been developed so far. 
By using such a degree-based ordering, 
the length of $v_i$'s adjacency list will tend to be smaller than 
that of $v_j$'s for every edge $(v_i, v_j)$ with $i < j$.  
Therefore, if $v_j$'s adjacency list is considerably larger than that of $v_i$'s, 
then it is beneficial to employ the $\langle j, i, k \rangle$ enumeration scheme 
and hash $v_j$'s adjacency list.
In our previous work on triangle counting algorithms for shared-memory 
platforms~\cite{tom2017exploring}, the $\langle j, i, k \rangle$ 
enumeration scheme along with map-based intersection approaches 
gave us faster runtimes as compared to other approaches.
In order to use the $\langle j, i, k \rangle$ 
enumeration scheme, the non-zero entries of $L$ are used and $C[L]$ is used 
to denote the non-zero entries of $L$. As in Equation~\ref{eq:7}, $C[L]$ is equal to $UL$.

\subsection{Cannon's parallel matrix multiplication algorithm}
Cannon's algorithm is a widely used method for multiplying two matrices in 
parallel~\cite{cannon1969cellular}. It is a dense, memory-efficient matrix multiplication 
algorithm that performs the multiplication operation in $\sqrt{p}$ shifts. 
Let $A$ and $B$ be the input matrices that need to be multiplied, 
and $C$, be the resulting matrix, i.e., $C=AB$.
The algorithm assumes a 2D block 
distribution of the input matrices and distributes these blocks across the square ($\sqrt{p} \times \sqrt{p}$) processor grid such 
that processor $P_{i,j}$ locally stores the blocks $A_{i, \small(j+i\small)\%\sqrt{p}}$ 
and  $B_{\small(i+j\small)\%\sqrt{p},  j}$. The algorithm multiplies these local blocks first. 
The local blocks of $A$ and $B$ are then shifted left along the row and up along the column, 
respectively, and multiplied again. For dense matrix multiplication, every processor is busy 
with computations after each shift, and the number of communications is bounded at $\sqrt{p}$. 
\vspace*{-2mm}

%% file: related.tex
\vspace*{-2mm}
\section{Related work}
\label{related}
Distributed memory algorithms are based on concepts similar to the ones described in Section~\ref{background}. 

Pearce~et~al.~\cite{pearce2017triangle} developed a triangle counting application over HavoqGT~\cite{6569865}, an asynchronous 
vertex-centric graph analytics framework for distributed-memory. Their approach 
includes distributing the undirected graph using \emph{distributed delegates}~\cite{pearce2017triangle}, 
followed by a two-core decomposition of the graph which removes the vertices that cannot be a 
part of any triangle. After this, they proceed with reordering their vertices based on degree 
to compute wedges. These wedges, partitioned using 1D decomposition, 
are henceforth queried for closure to check the existence 
of a triangle. 

Arifuzzaman~et~al.~\cite{arifuzzaman2017distributed} 
developed two different approaches, one which  
avoids communication with overlapping 
partitions whereas, the other which optimizes on memory usage. 
In the communication-avoiding approach, the vertices 
of the graph are partitioned into $p$ disjoint subsets, where, $p$ is the number of processors. 
Each processor is responsible for storing one of these subsets and their corresponding 
adjacency lists. Additionally, in order for each processor to work independently, 
the adjacency lists of the adjacent vertices of these vertices are stored too. 
Since most real-world sparse graphs follow a power-law degree distribution, 
a naive partitioning of the vertices of such a graph will lead to high memory overheads 
as the lengths of adjacency lists will be very skewed. 
Furthermore, the triangle counting operation will also incur high load imbalance, 
which will negatively impact the performance. Arifuzzaman~et~al. have explored these aspects 
and have developed various partitioning schemes in order to load balance their computations. 
In order to reduce the memory overheads of the above approach, Arifuzzaman~et~al. 
have further developed a space-efficient method, which involves higher communication costs. In this 
approach, they partitioned the vertices across processors into disjoint subsets and only stored the adjacency 
list of these vertices. Subsequently, only one copy of the graph exists across all the processors. 
For every intersection operation, they follow a push-based mechanism, in which the adjacency list 
of a vertex is sent to processors which require this particular list for performing the intersection. 
However, this leads to high communication overheads. 

Kanewala~et~al.~\cite{kanewala2018distributed} describes a distributed, shared-memory algorithm
to triangle counting. Their algorithm explores different combinations of the upper triangular part of the adjacency 
matrix and the lower triangular part of the adjacency matrix to perform the set intersection 
operations (Refer to Section~\ref{background}) between the adjacency lists. In order to parallelize the algorithm in a distributed setting, 
they perform a 1D decomposition of the adjacency matrix and send the adjacency list of a vertex to the rank which stores the adjacency lists of its adjacent vertex. However, in order to curb the number of messages generated, they block vertices and their adjacency lists and process them in blocks.

%% file: methods.tex
\section{Methods}
\label{methods}

This section presents our parallel algorithm for triangle counting in distributed memory parallel systems. 
Our algorithm reads the input graph, preprocesses it to reorder the vertices in 
non-decreasing degree among other operations, and stores the graph 
using compressed spare row (CSR) format prior to triangle counting. Our implementation is 
based on the map-based triangle counting approach outlined in~\cite{tom2017exploring}, 
which was briefly described in Section~\ref{background}. 

\subsection{Parallelization}
\label{ssec:num1}

\paragraph{Task Decomposition and Data Distribution}
In our algorithm, we treat the computations required for computing an entry of the 
$C[U]$ matrix (Equation~\ref{eq:6}) as an indivisible computational task.
We decompose the computations among the $p$ processors, by mapping the tasks associated with $C[U]$ using a 2D decomposition of $C$. Specifically, the processors are organized in the form of a $\sqrt{p}\times\sqrt{p}$ processor grid and each processor is responsible for the elements of $C[U]$ that exist in the $n/\sqrt{p} \times n/\sqrt{p}$ entries of $C$ that were assigned to it.

However, as we only consider the upper triangular matrix, 
a naive 2D block partitioning will lead to load imbalance. 
Moreover, as the vertices are sorted in non-decreasing degree, 
the length of the adjacency list increases as the vertex id 
increases. 
Therefore, the tasks associated with the extreme right and the lower part 
of the $C[U]$ matrix will be computationally more expensive as they 
employ such vertices for the intersection.
This further contributes to the load imbalance. 

To address both issues and evenly distribute the work between the processors,
we perform a cyclic distribution of $C$ over the 
$\sqrt{p}\times\sqrt{p}$ processor grid. 
%
Because of the degree-based ordering, successive rows/columns 
in the upper and lower triangular portions of the adjacency 
matrix will have similar number of non-zeros. Consequently, a 
cell-by-cell cyclic distribution 
will tend to assign a similar number of non-zeros (tasks) of $C[U]$ 
and at the same time, a similar number of light and heavy tasks to each processor.

%
Furthermore, in order to map the input blocks $U$ and $L$ 
with the tasks owned by the processors, 
we decompose the matrices using cyclic distribution over the processor grid.
After the decomposition, as the number of vertices assigned to a processor is 
not contiguous anymore, the adjacency list of a vertex $v_i$ is accessed using
the transformed index $v_i\div\sqrt{p}$ in the per-processor CSR representation.
Let the blocks $U_{x,y}$ and $L_{x,y}$ be the respective decomposition of $U$ and $L$
over the $\sqrt{p}\times\sqrt{p}$ grid
such that processor $P_{x,y}$ is responsible for those blocks.

\paragraph{Orchestration of Computation and Communication}
 
Consider the set of tasks $C[U_{x,y}]$ that processor $P_{x,y}$ is responsible for. 
For every task $(v_i, v_j)$ in that block, 
in order to apply Equation~\ref{eq:7}, it requires the adjacency lists of 
the set of $v_i$ and $v_j$ vertices. 
Thus, $P_{x,y}$ needs the blocks $U_{x,*}$ and $L_{*,y}$ to determine 
the number of triangles in $C[U_{x,y}]$, and following the convention of 
Equation~\ref{eq:7}, the associated computations can be written as
\begin{equation} \label{eq:4}
  C[U_{x,y}] = \sum\limits_{z} U_{x, z} L_{z, y}.
\end{equation}

Though this can be done by having each processor first collect the necessary rows and column blocks of matrices $U$ and $L$, respectively, and then proceed to perform the required computations, such an approach will increase the memory overhead of the algorithm. 
We address this problem by realizing that the above summation can be appropriately performed by  
the communication pattern of Cannon's 2D parallel matrix multiplication,
and we utilize the same pattern in our algorithm (refer Section~\ref{background}). 
In terms of this communication pattern, Equation~\ref{eq:2} can be rewritten as
(recall that $\%$ indicates a mod operation)
\begin{equation} \label{eq:5}
  C[U_{x,y}] = \sum\limits_{z=0}^{\sqrt{p}-1} U_{x,(x+y+z)\%\sqrt{p}} L_{{(x+y+z)\%{\sqrt{p}}},y}.
\end{equation}
\noindent
In each shift, for each non-zero element $(v_i, v_j)$ that exists in block $C[U_{x,y}]$,
we hash the row $v_i$ that exists in block $U_{x,(x+y+z)\%\sqrt{p}}$ 
and lookup the $v_k$ vertices that exist in the column $v_j$ in 
block $L_{{(x+y+z)\%{\sqrt{p}}},y}$ to find the number
 of triangles incident on the edge $(v_i, v_j)$. 

The initial shifts of Cannon's algorithm sends the block $U_{x,y}$ to
processor $(x, (y+x)\%\sqrt{p})$ and the block $L_{x,y}$ 
to processor $((x+y)\%\sqrt{p}, y)$. 
After performing the triangle counting operation on
the blocks associated with that shift,
the block $U_{x,y}$ is sent left to $P_{x,y-1}$ and the block of $L_{x,y}$ is
sent up over to $P_{x-1, y}$ in the next $\sqrt{p}-1$ shifts, and the triangle counting
operation is performed as before.
Every processor accumulates this count of triangles
corresponding to the tasks stored in $C[U_{x,y}]$
over the $\sqrt{p}$ shifts and, the same is globally reduced over all the
processors in the grid in the end. 

Finally, as discussed in Section~\ref{background}, 
to leverage the benefits of enumerating the triangles using the 
$\langle j, i, k \rangle$ scheme in the map-based triangle counting approach, 
we define a task based on the non-zero elements in $L$ 
instead of $U$, as $L$ contains the incidence list for each vertex 
$v_j$. 
Therefore, $L$, instead of $U$, is cyclically distributed to construct 
a task block, denoted by $C[L_{x,y}]$.

\subsection{Optimizations}
\label{opt}
We include several optimizations which leverage the characteristics of sparse graphs 
to further increase the performance of our distributed-memory algorithm. These are detailed below.  

\paragraph{Modifying the hashing routine for sparser vertices}
Due to the 2D decomposition and the fact 
that we perform the required computations by operating on 
blocks of $U$ and $L$, the lengths of the adjacency lists that are 
being intersected will tend to be smaller (on the average, they should be 
smaller by a factor of $\sqrt{p}$). A direct consequence of this is that even with a moderately sized hashmap, 
the number of collisions will tend to be smaller. In order to take advantage of this,
before we hash the adjacency of a vertex within the triangle counting routine, 
we heuristically determine if the vertex is involved in collisions by 
utilizing the length of the adjacency list of the vertex. If the length of the adjacency list 
is less than the size of the hash-map, then those vertices will face no collision while being hashed. 
Such vertices are hashed by performing a direct bitwise AND operation without involving any probing.

\paragraph{Doubly sparse traversal of the CSR structure}
As we are performing a block cyclic distribution of the upper and the lower triangular 
portion of the adjacency matrix, multiple vertices allocated to a processor may not contain 
any adjacent vertices. This is because each processor will have roughly $1/\sqrt{p}$ 
of the adjacency lists and if the vertices in the adjacency list has  
a degree in $U$ that is less than the degree of the vertex itself, 
then the adjacency list of that vertex could be rendered empty. However, these vertices 
can not be directly eliminated from the 
CSR structure due to the indexing scheme we use to avoid maintaining offsets. 
Therefore, in order to eliminate this unnecessary looping over these vertices
while performing the set intersection operation,  
we use a data structure that is inspired by the doubly compressed sparse 
row structure~\cite{bulucc2010highly} to store the task matrix, 
as well as the upper and the lower triangular portion of the 
adjacency matrix. 
In our algorithm, while creating the CSR structure for 
each processor, we also associate with it a list of vertices that contain 
non-empty adjacency lists. We use this list of vertices 
to index into the CSR structure; thus, we avoid any vertices
that have empty adjacency lists.

\paragraph{Eliminating unnecessary intersection operations}
While performing the intersection operation between the two adjacency lists, 
we only need to consider those triangle-closing vertices, $k$, that satisfy $k>j$.
However, as the adjacency lists are split over $\sqrt{p}$ processors, it is 
possible that some of the participating vertices are assigned to other processors, 
and performing an intersection operation with 
these entries will result in no triangles. 
Therefore, to overcome this problem, 
while performing the hashmap lookups,
we traverse row $v_i$ backwards, that is, from the last adjacent vertex of
$v_i$, and then break out of the loop as soon as we encounter an 
adjacent vertex id that is lesser than the last adjacent vertex id 
in column $v_j$. With this, we weed out all the unnecessary intersection 
operations that would otherwise result in no common vertices. 
Since the adjacency lists stored in the CSR structure 
are not required to be sorted for any given vertex, this optimization requires an 
initial sort before we start the triangle counting phase. However, the 
cost incurred by this sorting is amortized over the many set intersection 
operations that take place in this phase.

\paragraph{Reducing overheads associated with communication}
In order to eliminate the cost of serializing and deserializing memory during 
MPI communication steps, we allocate the memory associated with all of the 
information for a sparse matrix as a single blob, and ``allocate'' from within that blob the 
various arrays of the sparse matrix that are required for the processing of the algorithm. 
Specifically, we convert and store the blocks $U_{x,y}$ and $L_{x,y}$ 
as a blob of bytes before the shifts begin. This gives us some savings with respect 
to the amount of time spent in communication.

\subsection{Preprocessing}
Our algorithm assumes that the graph is initially stored 
using a 1D distribution, in which each processor has $n/p$ vertices 
and its associated adjacency lists. 
Given that, it proceeds to perform a series of preprocessing steps 
whose goals are to (i) perform a cyclic distribution of the graph 
followed by a relabeling of the vertices, (ii) reorder the vertices of the graph 
in non-decreasing degree, (iii) redistribute the graph based on the 2D cyclic distribution 
required by our parallel formulation, and, (iv) create the upper
and the lower triangular portion of the adjacency matrix. The rational behind these pre-processing steps and our corresponding parallel formations are described in the rest of this section.
\vspace*{-2mm}
\paragraph{Initial redistribution}
In some cases the initial distribution of the graph may be 
such that even though each processor had $n/p$ vertices, it 
may have adjacency lists with significant variation in the number of 
non-zeros elements. 
In order to reduce the load imbalance incurred while 
preprocessing the graph datasets that contain localized sets of 
highly dense vertices, we perform an initial cyclic distribution of the graph 
and relabel the vertices in the adjacency list accordingly. 
\vspace*{-2mm}
\paragraph{Reorder vertices based on non-decreasing degree}
Recall from the discussion in Section~\ref{background}, prior to 
triangle counting, the graph vertices are reordered in non-decreasing degree 
as this significantly reduces the time required to do triangle counting.
In order to achieve the same, the vertices of the graph and the vertices in their respective adjacency 
lists are relabeled based on its position after sorting it in non-decreasing degree.
To perform this relabeling efficiently, we use distributed counting sort. 
Note that a side effect of this reordering is on the utilization of the available locality. 
Hash-map based triangle counting routines could potentially make use of the 
locality obtained by processing the vertices in an order such that 
vertices with similar adjacency lists are processed together. 
However, although the degree-based ordering destroys this locality, 
the gains achieved by such a reordering results in 
faster runtimes and is proven in~\cite{arifuzzaman2017distributed}.
\vspace*{-2mm}
\paragraph{Create the upper and the lower triangular portion of the adjacency matrix}
Recall from Subsection \ref{ssec:num1} that the upper and the 
lower triangular portions of the adjacency matrix are used in order to perform the computations 
involved in the triangle counting phase. 
The algorithm first processes its local chunk of vertices and the
associated adjacency lists to determine the vertices that will
need to be distributed to remote nodes to form a  
2D cyclic distribution. 
Once each node has received its portion of the adjacency matrix, 
a single scan through the adjacency list of each vertex in the chunk
is performed to create the upper and the lower triangular portions.
To convert the matrix into these triangular portions, 
the degree of a vertex is compared with the degree of an adjacent vertex, and 
the adjacent vertex is placed in the upper portion 
if its degree is greater than that of the former. If not, 
the vertex is placed in the lower portion of the adjacency matrix.
 Moreover, since the vertices are reordered based on non-decreasing degree,
the global position of the vertices can be used to compare the degrees 
of the vertices. However, in many scenarios, the position of the 
adjacent vertex is not locally available. Thus, this 
requires us to perform a communication step with all nodes which adds to the 
overheads in the parallel algorithm.   

\subsection{Cost Analysis}
\label{analysis}
In order to analyze our algorithm, we derive the parallel time complexity 
for the computation and the communication steps in the pre-processing 
and the triangle counting phases of our algorithm. 
Let $n$ be the total number of vertices of the graph, $m$ be the 
total number of edges of the graph, and $p$ the total number of ranks used. 
Moreover, let $d_{avg}$ and $d_{max}$ be the average degree 
and the maximum degree of the graph, respectively. 
The computations in the pre-processing phase, as discussed above, involve
multiple scans over the chunk of the adjacency matrix owned by each rank which
amounts to a computation time of $m/p$. Moreover, the communication step
in the pre-processing phase requires an all-to-all 
personalized communication operation. 
Since we implemented this communication step using $p$ point-to-point
send and receive operations, its complexity is lower bounded by $p + m/p$.
The pre-processing phase also includes the distributed counting 
sort for relabeling the vertices in non-decreasing degree order. 
The computations associated with this sort includes two scans of the local 
vertices to determine the local maximum degree and the local positions,  
an overall reduction to compute the global maximum degree and, 
a scan with a cost of $d_{max}$ to 
determine the new labels of the vertices by computing the new positions
in the distributed system. Furthermore, to find 
the new positions, we perform a prefix sum which incurs a communication 
time of $d_{max}\log(p)$.
Thus, the total pre-processing phase takes time
\[ 
T_{pre-processing} = p + \frac{m}{p} + \frac{n}{p} + \log(p) + d_{max} + d_{max}\log(p).
\] 

In the triangle counting phase, for each shift, the amount of computations 
is on the average $(n/\sqrt{p}) \times (d^2_{avg}/p)$, whereas the 
communication cost is $(n/\sqrt{p}) \times (d_{avg}/\sqrt{p})$. As a result, 
the overall amount of time, across the $\sqrt{p}$ shifts, 
spent in computation and communication for the triangle counting phase is
\[
T_{triangle-counting} = d_{avg} \frac{n}{\sqrt{p}} \Bigg(\frac{d_{avg}}{\sqrt{p}} + 1\Bigg).
\]

%% file: experimental.tex
\begin{table}[t]
\centering
\begin{threeparttable}
\caption{Datasets used in the experiments.}
\label{datasets}
\begin{tabular}{@{\hspace{1em}}lrrr@{\hspace{1em}}}
\toprule
\midrule
  \multicolumn{1}{c}{Graph} 	& 
  \multicolumn{1}{c}{\#vertices} 	& 
  \multicolumn{1}{c}{\#edges} 	&  
  \multicolumn{1}{c}{\#triangles} \\ 
\midrule
 twitter~\cite{kwak2010twitter}		& 41,652,230	& 	1,202,513,046	& 34,824,916,864 \\
 friendster~\cite{graphchallenge}	& 119,432,957	& 	1,799,999,986	&        191,716 	\\
 g500-s26~\cite{nelson2014grappa} 	&  67,108,864  & 	1,073,741,824  & 49,158,464,716 \\
 g500-s27~\cite{nelson2014grappa} 	& 134,217,728  & 	2,147,483,648  & 106,858,898,940 \\
 g500-s28~\cite{nelson2014grappa} 	& 268,435,456  & 	4,294,967,296  & 231,425,307,324 \\
 g500-s29~\cite{nelson2014grappa} 	& 536,870,912  & 	8,589,934,592	& 499,542,556,876\\
\bottomrule
\end{tabular}
\small
Summary of the graphs that were used to evaluate the performance of our triangle
counting algorithms. The number of triangles involved in each graph is listed as
well. 
\end{threeparttable}
\vspace{-1em}
\end{table}


\input{scaling_nums}

\input{scaling_graphs}

\section{Experimental methodology}
\label{experimental}
\subsection{Datasets}
We used six large, real-world and synthetic graphs with varying degree
distributions to evaluate the performance of our algorithms. Various statistics
related to these graphs and the sources from where they were obtained are shown in
Table~\ref{datasets}. \emph{twitter} and \emph{friendster} datasets 
are sparse, social networks, \emph{g500-s26}, \emph{g500-s27}, 
\emph{g500-s28} and \emph{g500-s29} were generated 
using the \texttt{graph500} generator provided in~\cite{nelson2014grappa}. 
These follow the \emph{RMAT} graph specifications~\cite{graph500}.
Our algorithm creates these synthetic graphs as input to each run prior
to calling our triangle counting routine. This way, we avoid reading the 
big graphs from the disk. We converted all the graph datasets to 
undirected, simple graphs.

\subsection{Experimental setup}
Our experiments were conducted on up to 29 dual-socket nodes 
with 24 cores of Intel Haswell E5-2680v3, each with 64 GB memory. 
More details about the system is detailed in~\cite{mesabi}. 
Our programs were developed using C and OpenMPI (v3.1.2), and compiled using 
GCC (v8.1.0) with -O3 optimization. We ran our MPI programs with the option
\texttt{--bind-to core}. We ran our experiments with ranks ranging from 16 to 169, 
such that the number of ranks is a perfect square which forms a $\sqrt{p}\times\sqrt{p}$ processor grid. 
The number of nodes used for the different number of ranks is detailed in Table~\ref{table:scalingnums}. 
In order to best utilize the available resources, we resort to minimizing the 
maximum number of nodes used such that the aggregate memory that exists in 
the selected number of nodes satisfies the memory requirement of our algorithm. 
Moreover, we have bound cores to socket as well. For example, 
with 36 ranks, we bind one core per socket to get better performance
while ensuring our program does not run out of memory.    
We obtain runtimes starting with 16 ranks, as the largest
graph in our experiments, i.e., \emph{g500-s29}, required all the memory 
provided by 16 nodes for the processing. 

Our competing approach, \emph{Havoq}, was executed on 48 nodes, with all 24 cores 
being used across the node. The program was compiled using GCC (v5.4.0) and OpenMPI (v3.1.2).
We had to resort to this particular version of GCC since \emph{Havoq} ran successfully on 
this version. Moreover, \emph{Havoq}  produced faster runtimes on this version than on GCC (v8.1.0). 

\subsection{Performance metrics}
We make use of the following two performance metrics 
for assessing the performance of our parallel formulation
\begin{enumerate}
\vspace*{-2mm}
\item {\emph{Speedup} - Speedup is computed
by dividing the runtime obtained by the baseline algorithm against the runtime obtained by 
the parallel algorithm. We consider the runtime obtained 
with 16 ranks as the baseline and report the speedup of 
the algorithm computed against that.}
\item{\emph{Efficiency} - Similar to above, we compute the efficiency obtained by 
using the runtime of the 16-rank case as the baseline. 
Specifically, if $p$ is the number of processors and  $T_p$ is the parallel 
runtime of our algorithm (or one of its phases), then we use 
$16T_{16}/pT_{p}$.}
\end{enumerate}

\section{Results}

\subsection{Parallel Performance}

Table~\ref{table:scalingnums} shows the performance achieved by our algorithm
on the two larger synthetic graphs and the two real-world graphs.  
%
%
From these results, we notice that as the number of ranks increase, the time required by the pre-processing and triangle counting times decrease. The overall speedup on 169 ranks relative to 16 ranks is in the range of $3.06$ to $6.93$ (compared to an expected speedup of $10.56$).
Note that the synthetic graphs achieving better speedups over the real-world graphs 
is due to the fact that the synthetic graphs we experiment on are larger.
Since the performance advantage of our algorithm tapers off after a certain number of ranks (which is also contingent on the size of the dataset), we restrict ourselves to measuring runtimes up until $13\times13$ processor grid.

Comparing the scalability of the pre-processing and the triangle counting
phases, we can see that the latter scales better. The relative speedup of
the triangle counting phase on $169$ ranks is on the average $1.71$ times 
higher than that achieved by the pre-processing phase.  This can also be 
seen in the efficiency plots of Figure~\ref{fig:scaling_graphs}, where the 
efficiency of the pre-processing phase decreases faster than the 
triangle counting phase when the number of ranks increase. 
In light of the analysis presented in Section~\ref{analysis}, this scaling behavior was expected.
The communication and the computation of the pre-processing phase is of the same order.
However, the computation in the triangle counting phase is 
of the order $d_{avg}/\sqrt{p}$ more than the pre-processing phase. Thus, the 
triangle counting phase continues to scale better than the pre-processing phase 
with increasing number of ranks. 

Moreover, for almost all graphs, the performance at $25$ ranks shows 
a super-linear speedup when compared to the runtimes obtained at $16$ ranks. We believe this 
happens because both the triangle counting and the pre-processing phase utilize caches 
better as the aggregate amount of cache memory increases with increasing number of ranks. 
This is further quantified in Figure~\ref{fig:rates}, which plots the operation rate in kOps/second for both
the phases for \emph{g500-s29}. We can see that although the pre-processing phase 
continues to show higher operation rates with increasing ranks, the triangle counting 
phase shows its peak performance at $25$ ranks.

\begin{figure}[t]
      \label{fig:rates}
        \includegraphics[width=2.5in]{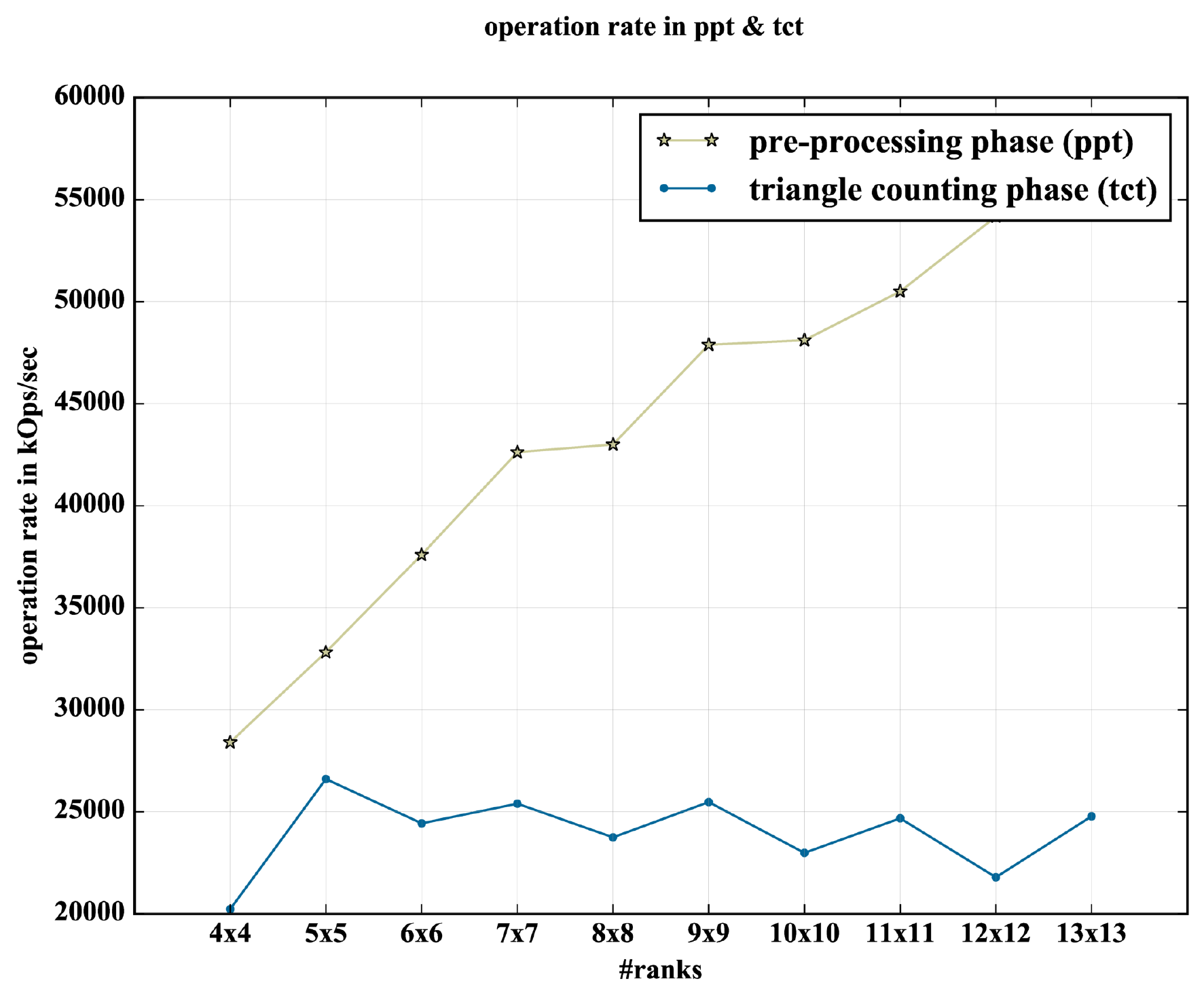}%
      \caption{This plot corresponds to the average operation rate in kOps per second achieved by the pre-processing phase and the triangle counting phase for \emph{g500-s29} over ranks from 16 to 169.}
      \label{fig:rates}
\end{figure}

Finally, on analyzing the real-world graphs, 
we notice that \emph{twitter} attains better speedups than \emph{friendster}. 
We believe this happens because the triangle counting phase in \emph{twitter} 
involves more work as opposed to \emph{friendster}. We measure the average 
number of probes performed per shift in every rank for a $169$ rank run 
in both \emph{twitter} and \emph{friendster}, and we observe that 
the number of probes in \emph{twitter} is $68$\% more than that of \emph{friendster}.
%

\subsection{Sources of overhead}

\begin{table}[t]
  \small
  \centering
  \begin{threeparttable}
    \caption{\emph{g500-s29} maximum runtime and load imbalance incurred per shift.}
    \label{table:r_l_imb}
    \begin{tabular}{lrrr}
      \toprule
      \midrule
      \multicolumn{1}{c}{ranks} 	& 
      \multicolumn{1}{c}{maximum runtime} 	& 
      \multicolumn{1}{c}{average runtime} 	&       
      \multicolumn{1}{c}{load imbalance} \\
      \midrule
	25	& 187.93	& 177.81 & 1.05\\
	36	& 106.65 	&  93.79 & 1.14\\
      \bottomrule
    \end{tabular}
    \begin{tablenotes}
    \footnotesize
        \item We measure the maximum runtime of a phase in the triangle counting routine and compute the associated load imbalance by dividing the maximum runtime per shift over average, in the \emph{g500-s29} dataset for 25 and 36 MPI ranks.
    \end{tablenotes}   
  \end{threeparttable}
\end{table}

We analyze three different sources of parallel overheads 
in our triangle counting algorithm. The first has to do with load imbalance as a 
result of assigning a different number of computations to each rank 
during each one of the $\sqrt{p}$ steps of the algorithm. 
The second is due to the redundant work
performed with increasing number of ranks. Finally, 
the third overhead we analyze is due to the time spent in communication as we
increase the number of ranks.
\vspace*{-2mm}
\paragraph{Load imbalance}
Recall from Section~\ref{methods}, that the computations during triangle counting is organized in $\sqrt{p}$ phases, and in each phase each processor performs a shift operation and processes the block of $U$ and $L$ that it just received. If the amount of computation associated with processing each pair of blocks is different, then this can lead to load imbalance. In order to quantify this load imbalance in the just the compute phase, we performed a series of experiments in which we measured the time obtained per shift for the computations involved in the triangle counting phase for 25 and 36 MPI ranks.
The load imbalance was measured as the ratio of the maximum amount of time over all pairs of blocks over the average time. The results are shown in Table~\ref{table:r_l_imb}.
For 25 MPI ranks, the load imbalance is $1.05$ and 
for 36 MPI ranks is $1.14$. 
We also quantify how distributing the data and the tasks contributes to 
the load imbalance. We count the number of non-zero tasks associated with
each rank with increasing grid sizes, and compute the load imbalance. 
In general, the load imbalance that we observed was less than $6\%$, which can further explain the load imbalance observed over $25$ and $36$ ranks runtimes.
\vspace*{-2mm}
\paragraph{Redundant work}
In Section~\ref{methods}, we discussed various optimizations that were designed to efficiently operate on the very sparse blocks of $U$ and $L$ (e.g., doubly sparse traversal), in order to eliminate redundant computations. However, those optimizations do not entirely eliminate the redundant computations. To measure how much extra work we do as we increase the number of ranks, we instrumented our code to count the 
number of tasks that result in the map-based set intersection operation throughout the execution of the triangle counting phase. The number of such tasks for \emph{g500-s29} on 16, 25, and 36 ranks are shown in Table~\ref{table:opcount}. 
We see that as the number of ranks grow from 16 to 25 and 25 to 36, 
the number of 
tasks increases by $25\%$ and $20\%$, respectively. 
This extra work is responsible for some of the loss in the potential speedup observed 
in the results shown in Table~\ref{table:r_l_imb}.

\begin{figure}[t]
      \label{fig:ppt_fraction}
        \includegraphics[width=2.5in]{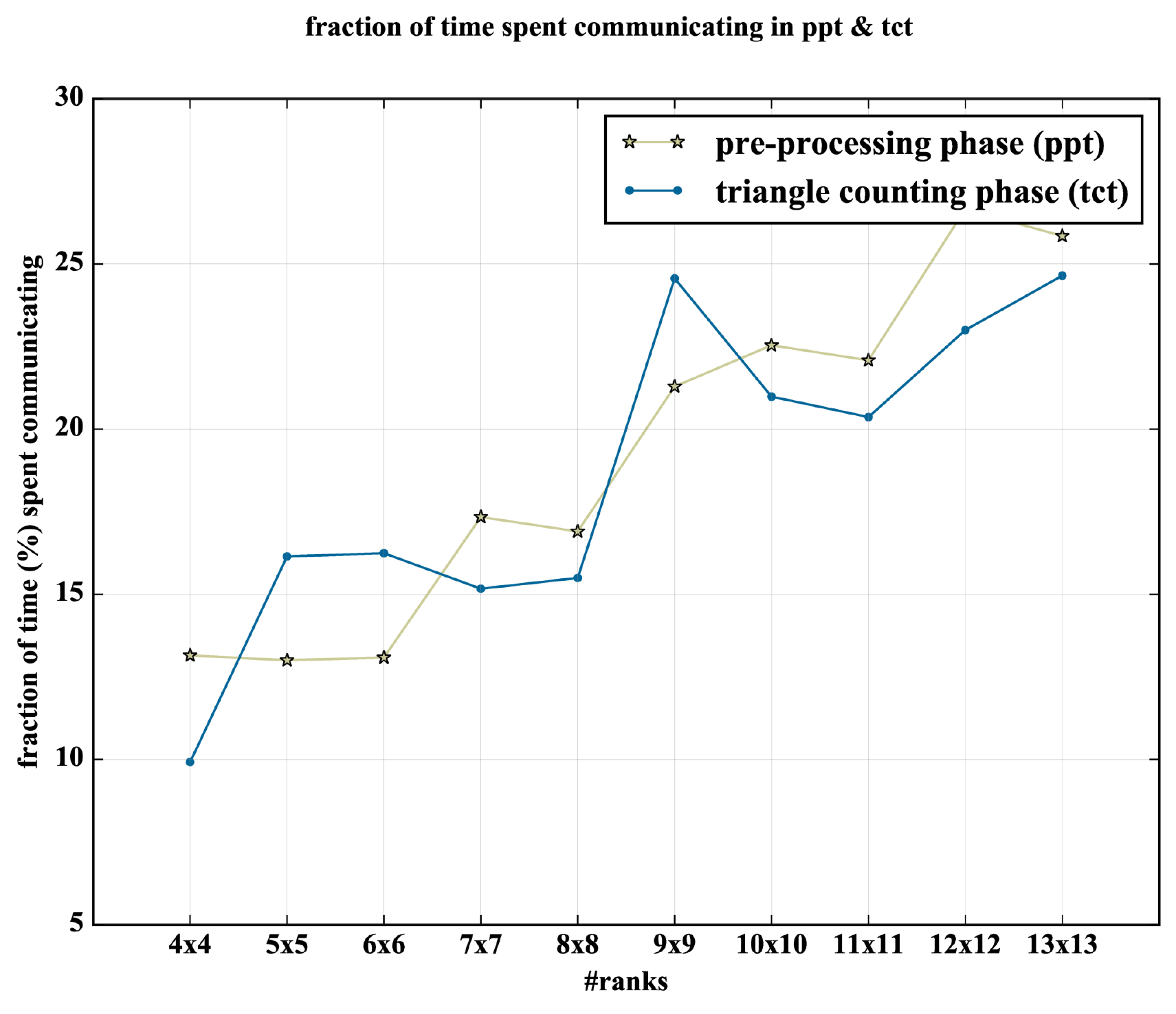}%
      \caption{This plot corresponds to the fraction of time (in percentage) taken by communication over the entire time taken by the pre-processing and the triangle counting phase for \emph{g500-s29}.}
      \label{fig:ppt_fraction}
\end{figure}

\paragraph{Communication overheads}

The fraction of time spent on communication over the entire runtime of the 
pre-processing and the triangle counting phase can be identified by looking at Figure~\ref{fig:ppt_fraction}. 
This plot shows that for both pre-processing and triangle counting, 
the bulk of the time is spent in computations for the largest graph in our testbed. 
However, the portion of the overall runtime 
that is attributed to the communication keeps increasing as we increase the number of ranks.

\subsection{Quantifying the gains achieved by the optimizations}
Recall from Section~\ref{opt} that we introduced various optimizations in our triangle counting 
phase that leverage the sparsity that occur in the graphs and the structure of the computations. In order
to quantify the reduction in the runtime of the triangle counting phase, we choose the first two optimizations, which we believe gave us maximum benefits: (i) using a doubly sparse traversal of the CSR structure, and, (ii) modifying the hashing routine for sparser vertices, and, recorded the runtime of the triangle counting phase without using these optimizations. 

Based on the results that we obtained from these experiments on \emph{g500-s29}, we observe that
the doubly-sparse traversal of the vertices has
reduced the runtime of the triangle counting phase by $10\%$ and $15\%$ 
for $16$ ranks and $100$ ranks, respectively. In a similar light, 
the modified hashing routine has reduced the runtime of the 
triangle counting phase by $1.2\%$ and $8.7\%$ for $16$ ranks and 
$100$ ranks, respectively. Moreover, we also recorded the improvement obtained by using the  
enumeration scheme as $\langle j, i, k \rangle$ as opposed to $\langle i, j, k \rangle$ in our 
algorithm. We observe that the triangle counting runtime decreased by $72.8\%$ when we 
use the $\langle j, i, k \rangle$ enumeration scheme as compared to the 
$\langle i, j, k \rangle$ enumeration scheme.

\subsection{Comparison against other algorithms}

\begin{table}[b]
  \small
  \centering
  \begin{threeparttable}
    \caption{\emph{g500-s29} task count growth with respect to the number of ranks.}
    \label{table:opcount}
    \begin{tabular}{rrc}
      \toprule
      \midrule
      \multicolumn{1}{c}{ranks} 	&  \multicolumn{1}{c}{task} & \multicolumn{1}{c}{percent increase with respect}\\
      \multicolumn{1}{c}{used} 	&  \multicolumn{1}{c}{counts} & \multicolumn{1}{c}{to previous rank}\\
      \midrule
	$16$	& $33907905131$ & \\
	$25$	& $42360246067$ & $25\%$\\
	$36$	& $50801950709$ & $20\%$\\
      \bottomrule
    \end{tabular}
    \begin{tablenotes}
    \footnotesize
        \item We count the number of tasks that result in the map-based 
        set intersection operation in the \emph{g500-s29} dataset to measure the 
        redundant work with increasing number of ranks.
    \end{tablenotes}   
  \end{threeparttable}
\end{table}

\input{havoq_comparison}

\input{comp_app}

\paragraph{Comparison against \emph{Havoq}~\cite{pearce2017triangle}}
As discussed in Section~\ref{related}, various distributed memory parallel triangle counting algorithms have been developed. We perform two different evaluations. First, we perform a direct comparison with \emph{Havoq} on the graphs detailed in Table~\ref{datasets}. Table~\ref{table:havoq_comparison} compares the triangle counting runtime obtained by \emph{Havoq} and the triangle counting time obtained by our approach. \emph{Havoq} runtimes were obtained on 1152 cores (using 48 nodes) and our runtimes were obtained on 169 cores. On average, we get a speedup of $10.2$ times over their approach. In \emph{friendster}, our approach is slower than \emph{Havoq}. We believe this is because \emph{Havoq} does an edge-based partitioning scheme (referred to as \emph{delegate partitioning}), which leads to better scaling capability as compared to our method, which incurs more overheads as the number of ranks increase. Furthermore, \emph{Havoq} required more number of nodes than what was available in our system for \emph{g500-s29} and we could not obtain the runtime for the same.
\vspace*{-2mm}
\paragraph{Comparison against other distributed-memory algorithm~\cite{arifuzzaman2017distributed, kanewala2018distributed}}
We also contrast the performance achieved by our algorithm against what was achieved by other previous approaches on only the \emph{twitter} graph, since this was the common benchmark. For this second evaluation, we use the runtimes that were reported in the respective papers, which were obtained on different architectures and number of ranks. Thus, the comparisons presented with these approaches in Table~\ref{table:comp_app} should be interpreted in view of this caveat.
The performance achieved by the various algorithms on \emph{twitter} is shown in Table~\ref{table:comp_app}. 
Algorithm with Overlapping Partitioning (AOP)~\cite{arifuzzaman2017distributed} was run on 200 cores. Their experimental setup included 64 computing nodes (QDR InfiniBand interconnect) with 16 processors (Sandy Bridge E5-2670, 2.6GHz) per node, has 4GB memory per processor, and uses the operating system CentOS Linux 6.
Surrogate~\cite{arifuzzaman2017distributed} was also run on 200 cores and their experimental setup was the same as that of AOP. 
OPT-PSP algorithm~\cite{kanewala2018distributed} was run on 2048 cores. The experimental setup 
for this algorithm included a Cray XC system which has 2 Broadwell 22-core Intel Xeon processors, 
and the scaling experiments used only up to 16 cores to uniformly double the problem size.
From these results we can see that our implementation is comparable to all previous approaches. Moreover, the relative performance advantage of our method still holds, when we account for the fact that some of the runtimes reported in Table~\ref{table:comp_app} use more number of cores than those used in our experiments.


%% file: scaling_nums.tex
\begin{table*}[t]
  \small
  \centering
  \begin{threeparttable}
    \caption{Parallel Performance achieved using 16 - 169 MPI ranks.}
    \label{table:scalingnums}
    \begin{tabular}{
    	l@{\hspace{15pt}}
	r@{\hspace{15pt}}
	r@{\hspace{15pt}}
	r@{\hspace{15pt}}
	r@{\hspace{5pt}}
	r@{\hspace{15pt}}
	r@{\hspace{15pt}}
	r@{\hspace{5pt}}
	r@{\hspace{15pt}}
	r@{\hspace{15pt}}
	r@{\hspace{5pt}}
	r@{\hspace{15pt}}
	r@{\hspace{15pt}}}
      \toprule
      \midrule
       	\multicolumn{1}{@{\hspace{1pt}}c@{\hspace{5pt}}}{}	& 	
	\multicolumn{1}{@{\hspace{1pt}}c@{\hspace{5pt}}}{}		& 
	\multicolumn{1}{@{\hspace{1pt}}c@{\hspace{5pt}}}{\#nodes} 	&
	\multicolumn{1}{@{\hspace{1pt}}c@{\hspace{5pt}}}{expected} 	& 
	\multicolumn{1}{@{\hspace{1pt}}c@{\hspace{5pt}}}{} 	& 
	\multicolumn{1}{@{\hspace{1pt}}c@{\hspace{5pt}}}{ppt} 	& 
	\multicolumn{1}{@{\hspace{1pt}}c@{\hspace{5pt}}}{ppt}	& 
	%
	\multicolumn{1}{@{\hspace{1pt}}c@{\hspace{5pt}}}{} 	& 
	\multicolumn{1}{@{\hspace{1pt}}c@{\hspace{5pt}}}{tct} 	& 
	\multicolumn{1}{@{\hspace{1pt}}c@{\hspace{5pt}}}{tct}    	& 
	%
	\multicolumn{1}{@{\hspace{1pt}}c@{\hspace{5pt}}}{} 	& 
	\multicolumn{1}{@{\hspace{1pt}}c@{\hspace{5pt}}}{overall} & 
	\multicolumn{1}{@{\hspace{1pt}}c@{\hspace{5pt}}}{overall} \\ 

	\multicolumn{1}{@{\hspace{1pt}}c@{\hspace{5pt}}}{datasets} 	&
	\multicolumn{1}{@{\hspace{1pt}}c@{\hspace{5pt}}}{ranks} 	&
	\multicolumn{1}{@{\hspace{1pt}}c@{\hspace{5pt}}}{used} 		& 
	\multicolumn{1}{@{\hspace{1pt}}c@{\hspace{5pt}}}{speedup} 	& 
	\multicolumn{1}{@{\hspace{1pt}}c@{\hspace{5pt}}}{} 	& 
	\multicolumn{1}{@{\hspace{1pt}}c@{\hspace{5pt}}}{time} 		& 
	\multicolumn{1}{@{\hspace{1pt}}c@{\hspace{5pt}}}{speedup} 	& 
	%
	\multicolumn{1}{@{\hspace{1pt}}c@{\hspace{5pt}}}{} 	& 
	\multicolumn{1}{@{\hspace{1pt}}c@{\hspace{5pt}}}{time} 		& 
	\multicolumn{1}{@{\hspace{1pt}}c@{\hspace{5pt}}}{speedup}  	&
	%
	\multicolumn{1}{@{\hspace{1pt}}c@{\hspace{5pt}}}{} 	& 
	\multicolumn{1}{@{\hspace{1pt}}c@{\hspace{5pt}}}{runtime}	& 
	\multicolumn{1}{@{\hspace{1pt}}c@{\hspace{5pt}}}{speedup} 	\\

      \midrule

	g500-s28 & 16 	&  8 		&   	  	& &  151.71 	&   	   & &  576.30 &          & & 728.23 &   \\
			& 25 &  7 		& 1.56 	& &  113.80 	& 1.33 & &  408.21 &  1.41 & & 522.31 &  1.39  \\
			& 36 &  9 		& 2.25 	& &  76.19 	& 1.99 & &  291.11 &  1.98 & & 367.47 &  1.98  \\
			& 49 &  9 		& 3.06 	& &  63.96 	& 2.37 & &  222.64 &  2.59 & & 286.78 &  2.54  \\
			& 64 &  7 		& 4.00 	& &  56.40 	& 2.69 & &  221.65 &  2.60 & & 278.26 &  2.62  \\
			& 81 &  9 		& 5.06 	& &  46.12 	& 3.29 & &  160.30 &  3.60 & & 206.61 &  3.52  \\
			& 100 & 10 	& 6.25 	& &  43.12 	& 3.52 & &  136.90 &  4.21 & & 180.18 &  4.04  \\
			& 121 & 11 	& 7.56 	& &  40.22 	& 3.77 & &  121.33 &  4.75 & & 161.67 &  4.50  \\
			& 144 & 18 	& 9.00 	& &  34.35 	& 4.42 & &  105.60 &  5.46 & & 140.05 &  5.20  \\
			& 169 & 17 	& 10.56 	& & 30.69 		& 4.94 & &  79.82   &  7.22 & &110.51  &  6.59  \\

      \midrule
	g500-s29	& 16		& 16	& 		& & 323.50	& 		& & 1371.75		&		& & 1695.57		& \\
			& 25		& 25	& 1.56	& & 158.90	& 2.04	& & 731.32 		& 1.88	& & 890.56		& 1.90 \\
			& 36		& 18	& 2.25	& & 163.41	& 1.98	& & 697.33		& 1.97	& & 861.21		& 1.97 \\
			& 49		& 17	& 3.06	& & 126.98	& 2.55	& & 510.08		& 2.69	& & 637.41		& 2.66 \\
			& 64		& 16	& 4.00	& & 100.59	& 3.22	& & 473.49	 	& 2.90	& & 574.34		& 2.95 \\
			& 81		& 17	& 5.06	& & 88.60		& 3.65	& & 386.65		& 3.55	& & 475.55		& 3.57 \\
			& 100	& 17	& 6.25	& & 72.13		& 4.48	& & 280.05		& 4.90	& & 352.36		& 4.81 \\
			& 121	& 21	& 7.56	& & 65.69		& 4.92	& & 250.37		& 5.48	& & 316.24		& 5.36 \\
			& 144	& 24	& 9.00	& & 63.06		& 5.13	& & 232.09		& 5.91	& & 295.33		& 5.74 \\
			& 169	& 29	& 10.56	& & 53.54		& 6.04	& & 191.16		& 7.18	& & 244.81		& 6.93 \\

      \midrule

	twitter 	&16 	& 2	&  	   	& & 60.76 &  		& & 109.46 &   		&  & 170.45 &  \\
		 	&25 	& 2	& 1.56 	& & 39.59 & 1.53 	& & 64.73 	& 1.69 	&  & 104.50 & 1.63  \\
			&36 	& 2	& 2.25 	& & 39.63 & 1.53 	& & 61.33 & 1.78 	&  & 101.17 & 1.68  \\
			&49 & 3	& 3.06 	& & 33.45 & 1.82 	& & 45.31 & 2.42 	&  & 79.04   & 2.16  \\
			&64 & 3	& 4.00 	& & 30.16 & 2.01 	& & 42.13 & 2.60 	&  & 72.48   & 2.35  \\	
			&81 & 4	& 5.06 	& & 29.08 & 2.09 	& & 30.46 & 3.59 	&  & 59.68   & 2.86  \\
			&100 &  5	& 6.25 	& & 32.74 & 1.86 	& & 30.81 & 3.55 	&  & 63.70   & 2.68  \\
			&121 &  6	& 7.56 	& & 32.64 & 1.86 	& & 24.75 & 4.42 	&  & 57.50   & 2.96 \\
			&144 &  7	& 9.00 	& & 33.36 & 1.82 	& & 25.39 & 4.31 	&  & 58.85   & 2.90 \\
			&169 &  8	& 10.56 	& & 31.62 & 1.92 	& & 18.52 & 5.91 	& & 50.29    & 3.39 \\
      \midrule

	friendster 	&16 	&  3 &  		& & 91.54 &  		& & 95.41 &  		& & 	187.21 &   \\
			& 25 &  2 & 1.56 	& & 57.84 & 1.58 	& & 71.82 & 1.33 	& & 	129.78 & 1.44 \\
			& 36 &  3 & 2.25 	& & 48.51 & 1.89 	& & 64.29 & 1.48 	& & 	112.98 & 1.66  \\
			& 49 &  5 & 3.06 	& & 36.47 & 2.51 	& & 46.75 & 2.04 	& & 	83.37 & 2.25  \\
			& 64 &  4 & 4.00 	& & 35.80 & 2.56 	& & 45.61 & 2.09 	& & 	81.66 & 2.29  \\
			& 81 &  5 & 5.06 	& & 33.24 & 2.75 	& & 35.36 & 2.70 	& & 	68.78 & 2.72 \\
			& 100 & 5 & 6.25	& & 35.56 & 2.57 	& & 35.24 & 2.71 	& & 	71.04 & 2.64 \\
			& 121 & 7 & 7.56 	& & 29.51 & 3.10 	& & 27.51 & 3.47 	& & 	57.09 & 3.28 \\
			& 144 & 6 & 9.00 	& & 38.62 & 2.37 	& & 37.65 & 2.53 	& & 	76.53 & 2.45  \\
			& 169 & 8 & 10.56 	& & 31.55 & 2.90 	& & 29.43 & 3.24 	& & 	61.23 & 3.06  \\
			
      \bottomrule
    \end{tabular}
    \small
    \begin{tablenotes}
    \item{
     The column labeled \emph{ppt} shows the preprocessing runtime, \emph{tct} shows the triangle counting runtime,
     and the column labeled \emph{overall} is the overall runtime for the datasets. The runtimes are in seconds. 
     The speedup and efficiency were computed relative to the 16 rank runtimes.     
     The column labeled \emph{\#nodes used} correspond to the number of nodes 
     used to run the algorithm by distributing the various ranks over it.
     We select the number of nodes to minimize the maximum number of nodes used such that the aggregate memory 
     that exists in the selected number of nodes satisfies the memory requirement of our algorithm.}
    \end{tablenotes}
  \end{threeparttable}
\end{table*}

%% file: scaling_graphs.tex
\begin{figure*}[t]
      \centering
      \subfloat[\label{fig:scale28}]{%
        \includegraphics[width=0.42\textwidth]{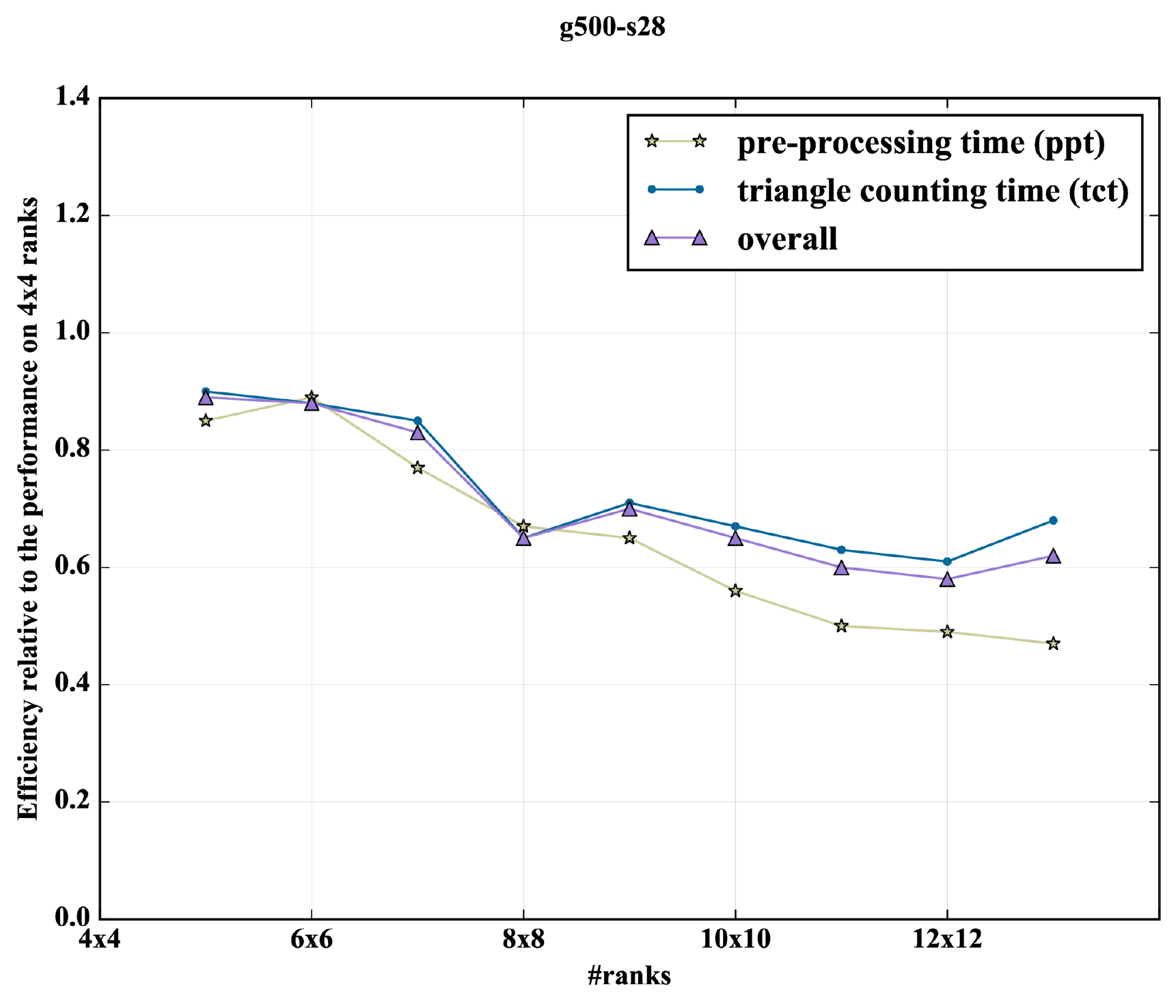}%
      }
      \hfill
      \subfloat[\label{fig:scale29}]{%
        \includegraphics[width=0.42\textwidth]{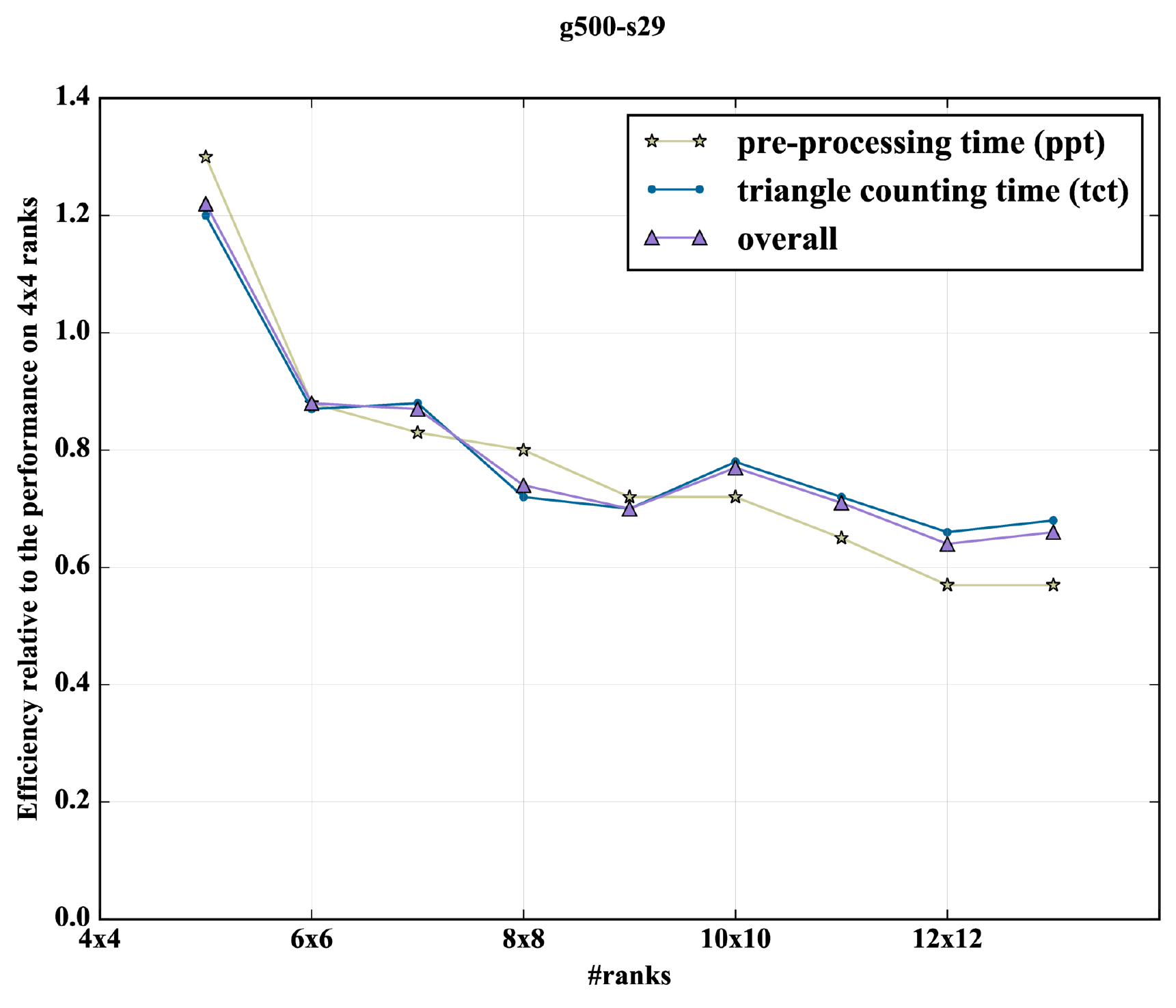}%
      }
      \hfill
      \subfloat[\label{fig:twitter}]{%
        \includegraphics[width=0.42\textwidth]{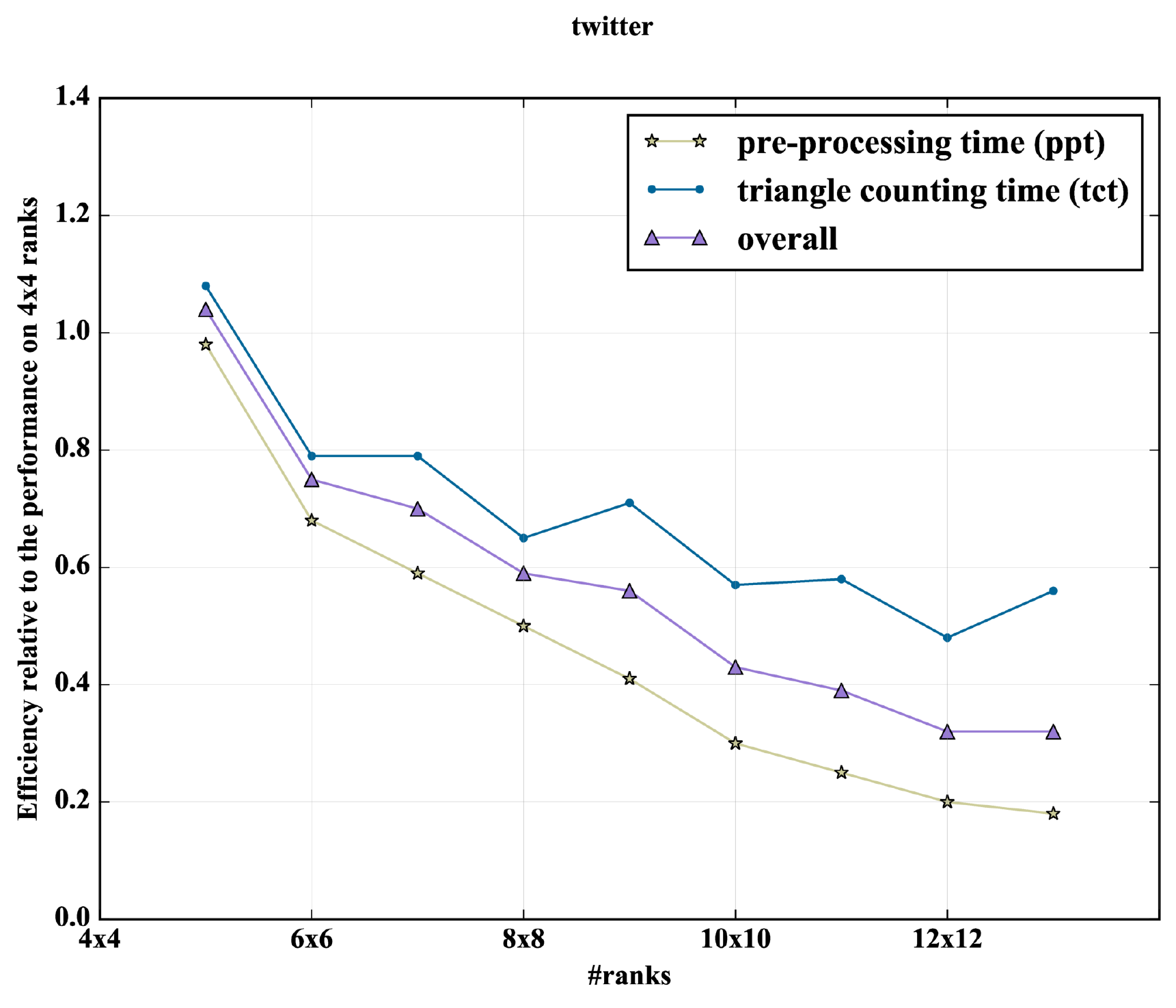}%
      }
      \hfill
      \subfloat[\label{fig:friendster}]{%
        \includegraphics[width=0.42\textwidth]{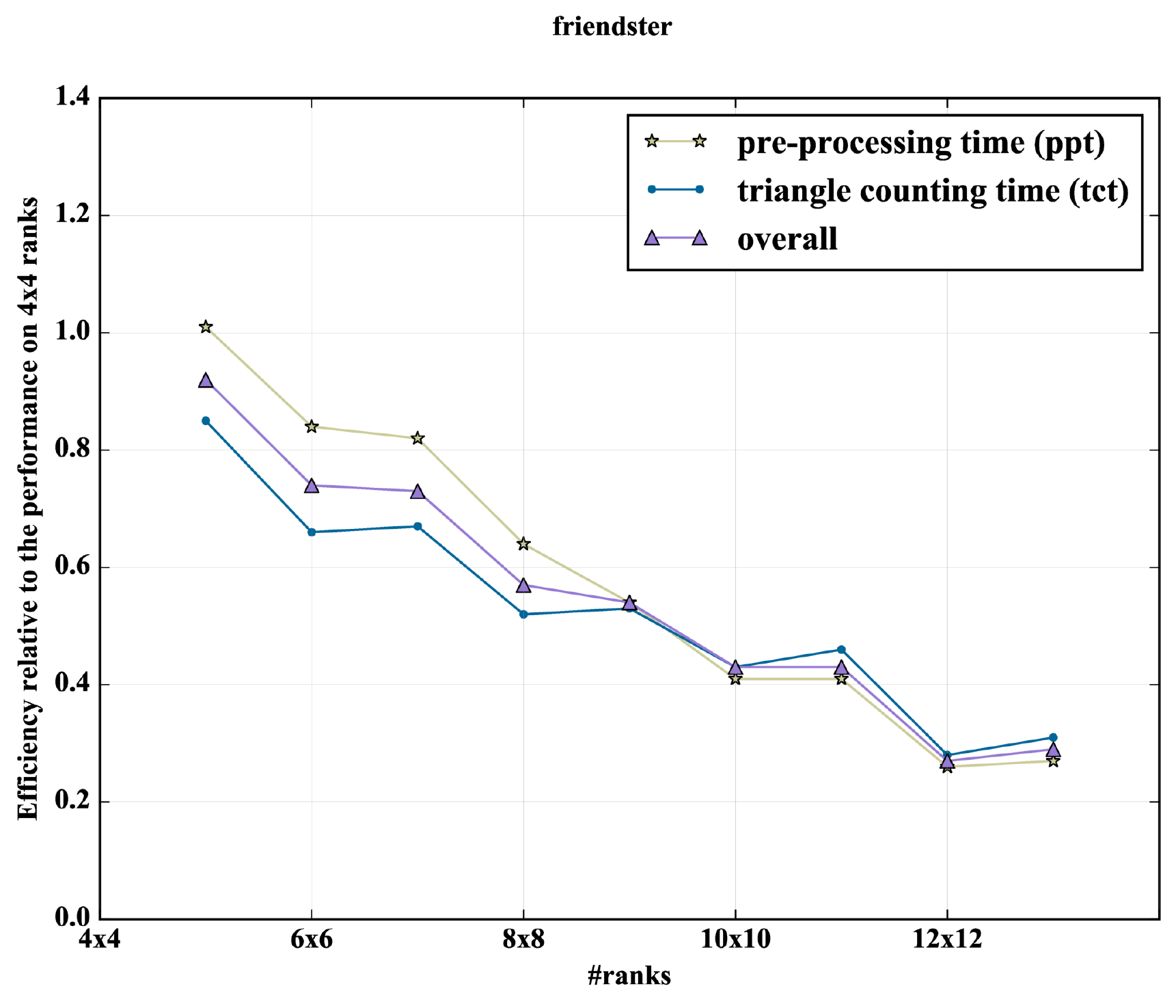}%
      }
      \caption{Efficiency achieved by our algorithm in the preprocessing step (labeled as ``ppt''), the triangle counting step (labeled as ``tct'') and the overall time (label as ``overall'') is plotted for the datasets while using the runtime obtained on the $4\times4$ processor grid as the baseline.}
      \label{fig:scaling_graphs}
\end{figure*}

%% file: havoq_comparison.tex
\begin{table}[h]
  \small
  \centering
  \begin{threeparttable}
    \caption{Comparisons with \emph{Havoq}'s~\cite{pearce2017triangle} triangle counting runtime}
    \label{table:havoq_comparison}
    \begin{tabular}{rrrrr}
      \toprule
      \midrule   
      \multicolumn{1}{c}{} & \multicolumn{1}{c}{2core} &  \multicolumn{1}{c}{directed wedge} & \multicolumn{1}{c}{our} & \multicolumn{1}{c}{speedup} \\
      \multicolumn{1}{c}{dataset} & \multicolumn{1}{c}{time}   & \multicolumn{1}{c}{counting time} & \multicolumn{1}{c}{runtime} & \multicolumn{1}{c}{obtained}\\
      \midrule
	g500-s26 	& 1.59	& 239.64	& 20.35 & 11.9 \\
	g500-s27 	& 3.37	& 576.45	& 41.93 & 13.7 \\
	g500-s28 	& 7.32	& 1395.11	& 79.82 & 14.6 \\
	twitter 	& 1.88	& 124.72	& 18.52 & 6.2   \\
	friendster 	& 3.29	& 24.75	& 29.43 & $-$   \\
      \bottomrule
    \end{tabular}
    \begin{tablenotes}
    \footnotesize
        \item Runtimes obtained by \emph{Havoq}'s triangle counting routine on the different input datasets. 
        \item \emph{2core time} corresponds to the amount of time taken by \emph{Havoq} to generate directed wedges. \emph{directed wedge counting time} corresponds to the amount of time taken by \emph{Havoq} to count the existence of the wedges generated.
        \item We use the \emph{ingest\_edgelist} executable provided in the \emph{Havoq} executable to convert the input data to their format and persist it in \texttt{/dev/shm}. The \emph{2core} time and the \emph{directed wedge counting} time that are reported by \emph{Havoq} are added to get the total triangle counting time. 
    \end{tablenotes}   
  \end{threeparttable}
\end{table}

%% file: comp_app.tex
\begin{table}[t]
  \small
  \centering
  \begin{threeparttable}
    \caption{Twitter graph runtime contrasted against other distributed-memory triangle counting approaches.}
    \label{table:comp_app}
    \begin{tabular}{lrr}
      \toprule
      \midrule
      \multicolumn{1}{c}{algorithm} 	& \multicolumn{1}{c}{fastest runtime reported} 	&  \multicolumn{1}{c}{cores used} \\
      \midrule
	Our work								& 51.7	& 169 	\\
	AOP~\cite{arifuzzaman2017distributed}		& 564.0	& 200	\\
	Surrogate~\cite{arifuzzaman2017distributed}	& 739.8	& 200	\\
	OPT-PSP~\cite{kanewala2018distributed}		& $23.14^a$	& 2048	\\
      \bottomrule
    \end{tabular}
    \begin{tablenotes}
    \footnotesize
        \item We make comparisons with three state-of-the-art approaches in~\cite{arifuzzaman2017distributed, kanewala2018distributed} on the \emph{twitter} dataset against our algorithm using the fastest runtimes (in seconds) reported. We also report the number of cores they used in obtaining the runtimes. $a$ has been extrapolated from the strong scaling results in~\cite{kanewala2018distributed}, using the speedup achieved and their fastest sequential runtime.  
	
	
    \end{tablenotes}   
  \end{threeparttable}
\end{table}

%% file: conclusion.tex
\section{Conclusion}
In this paper we presented a distributed memory formulation for triangle counting and evaluated its performance on real-word and synthetic graphs. Compared to prior parallel formulations, our formulation utilizes a 2D decomposition which increases the concurrency that it can exploit while reducing the overall communication overhead. The experimental results showed that these features lead to good scaling performance. Our analysis also identified areas that can benefit from further algorithmic and data structure improvements in order to better balance the work and reduce the amount of redundant computations. Moreover, we also note that this work can be easily extended to deal with rectangular processor grids using the \emph{SUMMA}~\cite{van1997summa} algorithm.

%% file: distr_triangle_2D.bbl

\begin{thebibliography}{25}


\ifx \showCODEN    \undefined \def \showCODEN     #1{\unskip}     \fi
\ifx \showDOI      \undefined \def \showDOI       #1{#1}\fi
\ifx \showISBNx    \undefined \def \showISBNx     #1{\unskip}     \fi
\ifx \showISBNxiii \undefined \def \showISBNxiii  #1{\unskip}     \fi
\ifx \showISSN     \undefined \def \showISSN      #1{\unskip}     \fi
\ifx \showLCCN     \undefined \def \showLCCN      #1{\unskip}     \fi
\ifx \shownote     \undefined \def \shownote      #1{#1}          \fi
\ifx \showarticletitle \undefined \def \showarticletitle #1{#1}   \fi
\ifx \showURL      \undefined \def \showURL       {\relax}        \fi
\providecommand\bibfield[2]{#2}
\providecommand\bibinfo[2]{#2}
\providecommand\natexlab[1]{#1}
\providecommand\showeprint[2][]{arXiv:#2}

\bibitem[\protect\citeauthoryear{Arifuzzaman, Khan, and Marathe}{Arifuzzaman
  et~al\mbox{.}}{2017}]%
        {arifuzzaman2017distributed}
\bibfield{author}{\bibinfo{person}{Shaikh Arifuzzaman}, \bibinfo{person}{Maleq
  Khan}, {and} \bibinfo{person}{Madhav Marathe}.}
  \bibinfo{year}{2017}\natexlab{}.
\newblock \showarticletitle{Distributed-Memory Parallel Algorithms for Counting
  and Listing Triangles in Big Graphs}.
\newblock \bibinfo{journal}{\emph{arXiv preprint arXiv:1706.05151}}
  (\bibinfo{year}{2017}).
\newblock


\bibitem[\protect\citeauthoryear{Bisson and Fatica}{Bisson and Fatica}{2018}]%
        {bisson2018update}
\bibfield{author}{\bibinfo{person}{Mauro Bisson} {and}
  \bibinfo{person}{Massimiliano Fatica}.} \bibinfo{year}{2018}\natexlab{}.
\newblock \showarticletitle{Update on Static Graph Challenge on GPU}. In
  \bibinfo{booktitle}{\emph{2018 IEEE High Performance extreme Computing
  Conference (HPEC)}}. IEEE, \bibinfo{pages}{1--8}.
\newblock


\bibitem[\protect\citeauthoryear{Bulu{\c{c}} and Gilbert}{Bulu{\c{c}} and
  Gilbert}{2010}]%
        {bulucc2010highly}
\bibfield{author}{\bibinfo{person}{Ayd{\i}n Bulu{\c{c}}} {and}
  \bibinfo{person}{John~R Gilbert}.} \bibinfo{year}{2010}\natexlab{}.
\newblock \showarticletitle{Highly parallel sparse matrix-matrix
  multiplication}.
\newblock \bibinfo{journal}{\emph{arXiv preprint arXiv:1006.2183}}
  (\bibinfo{year}{2010}).
\newblock


\bibitem[\protect\citeauthoryear{Cannon}{Cannon}{1969}]%
        {cannon1969cellular}
\bibfield{author}{\bibinfo{person}{Lynn~E Cannon}.}
  \bibinfo{year}{1969}\natexlab{}.
\newblock \bibinfo{booktitle}{\emph{A CELLULAR COMPUTER TO IMPLEMENT THE KALMAN
  FILTER ALGORITHM.}}
\newblock \bibinfo{type}{{T}echnical {R}eport}. \bibinfo{institution}{MONTANA
  STATE UNIV BOZEMAN ENGINEERING RESEARCH LABS}.
\newblock


\bibitem[\protect\citeauthoryear{Girvan and Newman}{Girvan and Newman}{2002}]%
        {girvan2002community}
\bibfield{author}{\bibinfo{person}{Michelle Girvan} {and}
  \bibinfo{person}{Mark~EJ Newman}.} \bibinfo{year}{2002}\natexlab{}.
\newblock \showarticletitle{Community structure in social and biological
  networks}.
\newblock \bibinfo{journal}{\emph{Proceedings of the national academy of
  sciences}} \bibinfo{volume}{99}, \bibinfo{number}{12} (\bibinfo{year}{2002}),
  \bibinfo{pages}{7821--7826}.
\newblock


\bibitem[\protect\citeauthoryear{Graph500}{Graph500}{2018}]%
        {graph500}
\bibfield{author}{\bibinfo{person}{Graph500}.} \bibinfo{year}{2018}\natexlab{}.
\newblock \bibinfo{title}{graph500}.
\newblock
\newblock
\urldef\tempurl%
\url{https://graph500.org}
\showURL{%
\tempurl}


\bibitem[\protect\citeauthoryear{Green, Yalamanchili, and Mungu{\'\i}a}{Green
  et~al\mbox{.}}{2014}]%
        {green2014fast}
\bibfield{author}{\bibinfo{person}{Oded Green}, \bibinfo{person}{Pavan
  Yalamanchili}, {and} \bibinfo{person}{Llu{\'\i}s-Miquel Mungu{\'\i}a}.}
  \bibinfo{year}{2014}\natexlab{}.
\newblock \showarticletitle{Fast triangle counting on the GPU}. In
  \bibinfo{booktitle}{\emph{Proceedings of the 4th Workshop on Irregular
  Applications: Architectures and Algorithms}}. IEEE Press,
  \bibinfo{pages}{1--8}.
\newblock


\bibitem[\protect\citeauthoryear{Hu, Liu, and Huang}{Hu et~al\mbox{.}}{2018}]%
        {hu2018high}
\bibfield{author}{\bibinfo{person}{Yang Hu}, \bibinfo{person}{Hang Liu}, {and}
  \bibinfo{person}{H~Howie Huang}.} \bibinfo{year}{2018}\natexlab{}.
\newblock \showarticletitle{High-Performance Triangle Counting on GPUs}. In
  \bibinfo{booktitle}{\emph{2018 IEEE High Performance extreme Computing
  Conference (HPEC)}}. IEEE, \bibinfo{pages}{1--5}.
\newblock


\bibitem[\protect\citeauthoryear{Institute}{Institute}{2018}]%
        {mesabi}
\bibfield{author}{\bibinfo{person}{Minnesota~Supercomputing Institute}.}
  \bibinfo{year}{2018}\natexlab{}.
\newblock \bibinfo{title}{Mesabi Description}.
\newblock
\newblock
\urldef\tempurl%
\url{https://www.msi.umn.edu/content/mesabi}
\showURL{%
\tempurl}


\bibitem[\protect\citeauthoryear{Kanewala, Zalewski, and Lumsdaine}{Kanewala
  et~al\mbox{.}}{2018}]%
        {kanewala2018distributed}
\bibfield{author}{\bibinfo{person}{Thejaka~Amila Kanewala},
  \bibinfo{person}{Marcin Zalewski}, {and} \bibinfo{person}{Andrew Lumsdaine}.}
  \bibinfo{year}{2018}\natexlab{}.
\newblock \showarticletitle{Distributed, Shared-Memory Parallel Triangle
  Counting}. In \bibinfo{booktitle}{\emph{Proceedings of the Platform for
  Advanced Scientific Computing Conference}}. ACM, \bibinfo{pages}{5}.
\newblock


\bibitem[\protect\citeauthoryear{Kwak, Lee, Park, and Moon}{Kwak
  et~al\mbox{.}}{2010}]%
        {kwak2010twitter}
\bibfield{author}{\bibinfo{person}{Haewoon Kwak}, \bibinfo{person}{Changhyun
  Lee}, \bibinfo{person}{Hosung Park}, {and} \bibinfo{person}{Sue Moon}.}
  \bibinfo{year}{2010}\natexlab{}.
\newblock \showarticletitle{What is Twitter, a social network or a news
  media?}. In \bibinfo{booktitle}{\emph{Proceedings of the 19th international
  conference on World wide web}}. ACM, \bibinfo{pages}{591--600}.
\newblock


\bibitem[\protect\citeauthoryear{Nelson, Holt, Myers, Briggs, Ceze, Kahan, and
  Oskin}{Nelson et~al\mbox{.}}{2014}]%
        {nelson2014grappa}
\bibfield{author}{\bibinfo{person}{Jacob Nelson}, \bibinfo{person}{Brandon
  Holt}, \bibinfo{person}{Brandon Myers}, \bibinfo{person}{Preston Briggs},
  \bibinfo{person}{Luis Ceze}, \bibinfo{person}{Simon Kahan}, {and}
  \bibinfo{person}{Mark Oskin}.} \bibinfo{year}{2014}\natexlab{}.
\newblock \showarticletitle{Grappa: A latency-tolerant runtime for large-scale
  irregular applications}. In \bibinfo{booktitle}{\emph{International Workshop
  on Rack-Scale Computing (WRSC w/EuroSys)}}.
\newblock


\bibitem[\protect\citeauthoryear{Parimalarangan, Slota, and
  Madduri}{Parimalarangan et~al\mbox{.}}{2017}]%
        {parimalarangan2016fastpaper}
\bibfield{author}{\bibinfo{person}{Sindhuja Parimalarangan},
  \bibinfo{person}{George~M Slota}, {and} \bibinfo{person}{Kamesh Madduri}.}
  \bibinfo{year}{2017}\natexlab{}.
\newblock \showarticletitle{Fast Parallel Triad Census and Triangle Listing on
  Shared-Memory Platforms}. In \bibinfo{booktitle}{\emph{Parallel and
  Distributed Processing Symposium Workshop (IPDPSW), 2017 IEEE
  International}}. IEEE.
\newblock


\bibitem[\protect\citeauthoryear{Pearce}{Pearce}{2017}]%
        {pearce2017triangle}
\bibfield{author}{\bibinfo{person}{Roger Pearce}.}
  \bibinfo{year}{2017}\natexlab{}.
\newblock \showarticletitle{Triangle counting for scale-free graphs at scale in
  distributed memory}. In \bibinfo{booktitle}{\emph{2017 IEEE High Performance
  Extreme Computing Conference (HPEC)}}. IEEE, \bibinfo{pages}{1--4}.
\newblock


\bibitem[\protect\citeauthoryear{Pearce, Gokhale, and Amato}{Pearce
  et~al\mbox{.}}{2013}]%
        {6569865}
\bibfield{author}{\bibinfo{person}{R. Pearce}, \bibinfo{person}{M. Gokhale},
  {and} \bibinfo{person}{N.~M. Amato}.} \bibinfo{year}{2013}\natexlab{}.
\newblock \showarticletitle{Scaling Techniques for Massive Scale-Free Graphs in
  Distributed (External) Memory}. In \bibinfo{booktitle}{\emph{2013 IEEE 27th
  International Symposium on Parallel and Distributed Processing}}.
  \bibinfo{pages}{825--836}.
\newblock
\showISSN{1530-2075}
\urldef\tempurl%
\url{https://doi.org/10.1109/IPDPS.2013.72}
\showDOI{\tempurl}


\bibitem[\protect\citeauthoryear{Polak}{Polak}{2016}]%
        {polak2016counting}
\bibfield{author}{\bibinfo{person}{Adam Polak}.}
  \bibinfo{year}{2016}\natexlab{}.
\newblock \showarticletitle{Counting triangles in large graphs on GPU}. In
  \bibinfo{booktitle}{\emph{Parallel and Distributed Processing Symposium
  Workshops, 2016 IEEE International}}. IEEE, \bibinfo{pages}{740--746}.
\newblock


\bibitem[\protect\citeauthoryear{Samsi, Gadepally, Hurley, Jones, Kao,
  Mohindra, Monticciolo, Reuther, Smith, Song, Staheli, and Kepner}{Samsi
  et~al\mbox{.}}{2017}]%
        {graphchallenge}
\bibfield{author}{\bibinfo{person}{Siddharth Samsi}, \bibinfo{person}{Vijay
  Gadepally}, \bibinfo{person}{Michael Hurley}, \bibinfo{person}{Michael
  Jones}, \bibinfo{person}{Edward Kao}, \bibinfo{person}{Sanjeev Mohindra},
  \bibinfo{person}{Paul Monticciolo}, \bibinfo{person}{Albert Reuther},
  \bibinfo{person}{Steven Smith}, \bibinfo{person}{William Song},
  \bibinfo{person}{Diane Staheli}, {and} \bibinfo{person}{Jeremy Kepner}.}
  \bibinfo{year}{2017}\natexlab{}.
\newblock \showarticletitle{Static Graph Challenge: Subgraph Isomorphism}.
\newblock \bibinfo{journal}{\emph{IEEE HPEC}} (\bibinfo{year}{2017}).
\newblock


\bibitem[\protect\citeauthoryear{Shrivastava, Majumder, and
  Rastogi}{Shrivastava et~al\mbox{.}}{2008}]%
        {shrivastava2008mining}
\bibfield{author}{\bibinfo{person}{Nisheeth Shrivastava},
  \bibinfo{person}{Anirban Majumder}, {and} \bibinfo{person}{Rajeev Rastogi}.}
  \bibinfo{year}{2008}\natexlab{}.
\newblock \showarticletitle{Mining (social) network graphs to detect random
  link attacks}. In \bibinfo{booktitle}{\emph{Data Engineering, 2008. ICDE
  2008. IEEE 24th International Conference on}}. IEEE,
  \bibinfo{pages}{486--495}.
\newblock


\bibitem[\protect\citeauthoryear{Shun and Tangwongsan}{Shun and
  Tangwongsan}{2015}]%
        {shun2015multicore}
\bibfield{author}{\bibinfo{person}{Julian Shun} {and} \bibinfo{person}{Kanat
  Tangwongsan}.} \bibinfo{year}{2015}\natexlab{}.
\newblock \showarticletitle{Multicore triangle computations without tuning}. In
  \bibinfo{booktitle}{\emph{Data Engineering (ICDE), 2015 IEEE 31st
  International Conference on}}. IEEE, \bibinfo{pages}{149--160}.
\newblock


\bibitem[\protect\citeauthoryear{Smith, Liu, Ahmed, Tom, Petrini, and
  Karypis}{Smith et~al\mbox{.}}{2017}]%
        {smith2017truss}
\bibfield{author}{\bibinfo{person}{Shaden Smith}, \bibinfo{person}{Xing Liu},
  \bibinfo{person}{Nesreen~K Ahmed}, \bibinfo{person}{Ancy~Sarah Tom},
  \bibinfo{person}{Fabrizio Petrini}, {and} \bibinfo{person}{George Karypis}.}
  \bibinfo{year}{2017}\natexlab{}.
\newblock \showarticletitle{Truss decomposition on shared-memory parallel
  systems}. In \bibinfo{booktitle}{\emph{High Performance Extreme Computing
  Conference (HPEC)}}. IEEE, \bibinfo{pages}{1--6}.
\newblock


\bibitem[\protect\citeauthoryear{Tom, Sundaram, Ahmed, Smith, Eyerman,
  Kodiyath, Hur, Petrini, and Karypis}{Tom et~al\mbox{.}}{2017}]%
        {tom2017exploring}
\bibfield{author}{\bibinfo{person}{Ancy~Sarah Tom}, \bibinfo{person}{Narayanan
  Sundaram}, \bibinfo{person}{Nesreen~K Ahmed}, \bibinfo{person}{Shaden Smith},
  \bibinfo{person}{Stijn Eyerman}, \bibinfo{person}{Midhunchandra Kodiyath},
  \bibinfo{person}{Ibrahim Hur}, \bibinfo{person}{Fabrizio Petrini}, {and}
  \bibinfo{person}{George Karypis}.} \bibinfo{year}{2017}\natexlab{}.
\newblock \showarticletitle{Exploring optimizations on shared-memory platforms
  for parallel triangle counting algorithms}. In \bibinfo{booktitle}{\emph{High
  Performance Extreme Computing Conference (HPEC), 2017 IEEE}}. IEEE,
  \bibinfo{pages}{1--7}.
\newblock


\bibitem[\protect\citeauthoryear{Van De~Geijn and Watts}{Van De~Geijn and
  Watts}{1997}]%
        {van1997summa}
\bibfield{author}{\bibinfo{person}{Robert~A Van De~Geijn} {and}
  \bibinfo{person}{Jerrell Watts}.} \bibinfo{year}{1997}\natexlab{}.
\newblock \showarticletitle{SUMMA: Scalable universal matrix multiplication
  algorithm}.
\newblock \bibinfo{journal}{\emph{Concurrency: Practice and Experience}}
  \bibinfo{volume}{9}, \bibinfo{number}{4} (\bibinfo{year}{1997}),
  \bibinfo{pages}{255--274}.
\newblock


\bibitem[\protect\citeauthoryear{Wang, Wang, Yang, and Owens}{Wang
  et~al\mbox{.}}{2016}]%
        {wang2016comparative}
\bibfield{author}{\bibinfo{person}{Leyuan Wang}, \bibinfo{person}{Yangzihao
  Wang}, \bibinfo{person}{Carl Yang}, {and} \bibinfo{person}{John~D Owens}.}
  \bibinfo{year}{2016}\natexlab{}.
\newblock \showarticletitle{A comparative study on exact triangle counting
  algorithms on the GPU}. In \bibinfo{booktitle}{\emph{Proceedings of the ACM
  Workshop on High Performance Graph Processing}}. ACM, \bibinfo{pages}{1--8}.
\newblock


\bibitem[\protect\citeauthoryear{Watts and Strogatz}{Watts and
  Strogatz}{1998}]%
        {watts1998collective}
\bibfield{author}{\bibinfo{person}{Duncan~J Watts} {and}
  \bibinfo{person}{Steven~H Strogatz}.} \bibinfo{year}{1998}\natexlab{}.
\newblock \showarticletitle{Collective dynamics of small-world networks}.
\newblock \bibinfo{journal}{\emph{nature}} \bibinfo{volume}{393},
  \bibinfo{number}{6684} (\bibinfo{year}{1998}), \bibinfo{pages}{440--442}.
\newblock


\bibitem[\protect\citeauthoryear{Ya{\c{s}}ar, Rajamanickam, Wolf, Berry, and
  {\c{C}}ataly{\"u}rek}{Ya{\c{s}}ar et~al\mbox{.}}{2018}]%
        {yacsar2018fast}
\bibfield{author}{\bibinfo{person}{Abdurrahman Ya{\c{s}}ar},
  \bibinfo{person}{Sivasankaran Rajamanickam}, \bibinfo{person}{Michael Wolf},
  \bibinfo{person}{Jonathan Berry}, {and} \bibinfo{person}{{\"U}mit~V
  {\c{C}}ataly{\"u}rek}.} \bibinfo{year}{2018}\natexlab{}.
\newblock \showarticletitle{Fast Triangle Counting Using Cilk}. In
  \bibinfo{booktitle}{\emph{2018 IEEE High Performance extreme Computing
  Conference (HPEC)}}. IEEE, \bibinfo{pages}{1--7}.
\newblock


\end{thebibliography}
